\documentclass[journal=jacsat,manuscript=article]{achemso}

\usepackage[version=3]{mhchem} 
\usepackage{amsfonts}
\usepackage{hyperref}
\usepackage{cleveref}
\usepackage{xr-hyper}
\usepackage{color,soul}
\usepackage{changes}

\author{Nicolas H. Wong\textsuperscript{a}}
\affiliation[Georgia Tech]{\textsuperscript{\text{a}}Department of Chemical Engineering, Georgia Institute of Technology, Atlanta, GA 30363, United States}
\author{Julia H. Yang\textsuperscript{a}}
\affiliation[Georgia Tech]{\textsuperscript{\text{a}}Department of Chemical Engineering, Georgia Institute of Technology, Atlanta, GA 30363, United States}
\email{jhyang@gatech.edu}

\title[Fine-tuning MACE]
  {Bias in Universal Machine-Learned Interatomic Potentials and its Effects on Fine-Tuning}

\abbreviations{}

\begin{document}
\externaldocument{SI}
\newif\ifrevision

\ifrevision
  \newcommand{\add}[1]{\hl{#1}}
  \newcommand{\del}[1]{\textcolor{red}{\st{#1}}}
  \newcommand{\rep}[2]{\textcolor{red}{\st{#2}}\,\hl{#1}}
\else
  \newcommand{\add}[1]{#1}
  \newcommand{\del}[1]{}
  \newcommand{\rep}[2]{#1}
\fi







\begin{abstract}
Universal machine learned interatomic potentials (uMLIPs) embody a growing area of interest due to their transferability across the periodic table, displaying an error of about 0.6 kcal/mol against the Matbench Discovery test set. However, we show that achieving more accurate predictions on out-of-domain tasks requires fine-tuning. Additionally, we investigate the existence and influence of model biases in molecular dynamics (MD) by examining two approaches for data generation: from multiple MD trajectories in parallel, which we call naive fine-tuning, and from a single MD trajectory with fine-tuning after set intervals, which we call \rep{iterative}{periodic} fine-tuning. Our results find that naive fine-tuning generates constrained datasets that fail to represent MD simulations, and thus downstream fine-tuned models fail during extrapolation. In contrast, \rep{iterative}{periodic} fine-tuning yields models which are more generalizable and accurate, producing \rep{stable}{low-error} dynamics. These findings indicate the role of uMLIP bias in fine-tuning, and highlights the need for multiple fine-tuning steps. Lastly, we relate unphysical behavior to principal component space, and quantify extrapolations through Q-residual analysis, which are useful as a proxy for epistemic uncertainty for larger simulations.
\end{abstract}

\section*{Introduction}

Machine-learned interatomic potentials (MLIPs) model interatomic potentials by training on reference quantum mechanical calculations \cite{unke_machine_2021}. This data-driven approach is desirable \rep{for}{because of} its flexibility and generalizability to new chemistries, but requires more data and is more expensive than classical potentials \cite{rupp_fast_2012, vanommeslaeghe_charmm_2010, jorgensen_opls_1988}.
MLIPs have been rapidly developed over the past two decades: Early models were based on neural networks \cite{lorenz_representing_2004, behler_generalized_2007, behler_four_2021} and kernel regression methods \cite{bartok_gaussian_2010}. These methods used descriptors, such as smooth-overlap-of-atomic (SOAP) descriptors, to represent chemical environments. Unfortunately, these descriptors have strict functional forms\del{,} and thus have poor transferability to new chemistries \cite{behler_four_2021}. With the advent of deep learning and data-driven modeling, descriptors could be learned from data \cite{batzner_e3-equivariant_2022, batatia_design_2025}.
Because of this, MLIPs have evolved into graph neural networks \cite{gilmer_neural_nodate, batzner_e3-equivariant_2022, musaelian_learning_2023}, and more recently, transformers \cite{liao_equiformer_2023, liao_equiformerv2_2024, xie_gptff_2024}, which allow these models to be highly data efficient and transferable across chemistries. MLIPs have fundamentally transformed the materials science and chemistry fields by enabling evaluations of atomic structures approaching ab initio accuracy, which in turn have increased the length and time scale of quantum chemically accurate molecular dynamics (MD) simulations and reduced the cost of high-throughput screening \cite{wang_machine_2024, gubaev_accelerating_2019, elena_machine_2025, lee_accelerating_2025, hodapp_machine-learning_2021}. 

Lately, the community has seen rapid development of universal machine\add{-}learned interatomic potentials (uMLIPs) which utilize deep learning models to learn from vast datasets spanning chemistries such as inorganic materials \cite{deng_chgnet_2023, sriram_open_2024}, organic materials \cite{eastman_spice_2023}, and molecular systems \cite{levine_open_2025}. Notable models such as ORB \cite{neumann_orb_2024}, MACE \cite{batatia_foundation_2024}, CHGNet \cite{deng_chgnet_2023}, UMA \cite{wood_uma_2025}, and NequIP \cite{batzner_e3-equivariant_2022} against independently generated benchmarking datasets, such as Matbench Discovery \cite{riebesell_matbench_2024}, demonstrate errors of less than 0.1 eV/atom \cite{riebesell_matbench_2024, mazitov_pet-mad_2025}, demonstrating their high accuracy and transferability across different chemistries. For this reason, uMLIPs have gained interest due to their ability to approximate DFT-level accuracy \cite{gao_benchmarking_2025} for structural evaluations out of the box, making ab-initio-quality data more accessible across disciplines \cite{jacobs_practical_2025}. They have been used to simulate systems \del{and accelerate} such as lithium-ion battery electrolytes \cite{ju_application_2025, guo_machine_2024, hajibabaei_universal_2021}, predict material properties \cite{kholtobina_exploring_2025, gao_benchmarking_2025}, or sample MD trajectories \cite{gonzales_benchmarking_nodate}. However, as promising as this technology is, uMLIPs still exhibit extrapolatory behavior \cite{kreiman_understanding_2026}. 

Specifically, out-of-domain evaluations manifest as a systematic softening of the potential energy surface \cite{deng_systematic_2025} (PES). The PES represents a high-dimensional surface that captures interatomic interactions which govern molecular mechanics \cite{behler_generalized_2007}. Machine-learned potentials seek to emulate the PES using deep learning models trained on quantum mechanical data \cite{friederich_machine-learned_2021}. A systematic softening of the PES corresponds to a systematic underprediction, or bias, of forces and potential energies. 

This observation, among others \cite{liu_discrepancies_2023, loewUniversalMachineLearning2025b, ma_pretrained_2026}, is a result of domain shift, or covariate shift, and is a fundamental problem throughout deep learning \cite{wang_exposure_2020}, which arises when models are applied to domains outside their training distribution. This leads to errors as the models are forced to extrapolate beyond their training domain \cite{martius_extrapolation_2016}. This challenge is exacerbated by the high dimensionality of the models, which may capture spurious correlations in training data \cite{ye_clever_2025, sagawa_investigation_2020}. Such problems with generalization have been well studied throughout different applications\add{,} such as computer vision \cite{schneider_improving_2020, venkateswara_deep-learning_2017} and natural language processing \cite{elsahar_annotate_2019, zeng_explaining_nodate}\add{,} and only somewhat explored in materials science and chemistry \cite{kreiman_understanding_2026, klarner_drug_2023, cui_online_2025}. Thus, generalization remains a \rep{critical}{critial} limitation of MLIPs, as evaluating on structures at different temperatures \cite{edwards_exploring_2025}, pressures \cite{loewUniversalMachineLearning2025b}, compositions \cite{zhu_accelerating_2025}, and chemistries \cite{ju_application_2025} outside of the initial training dataset results in significant errors in predictions. 

In this work, we surmise that because of the high-dimensionality of chemical space, system-specific data will almost certainly not exist in the vast chemical datasets used to fit uMLIPs and result in limitations in transferability, which remain to be quantified. For example, the recently released Open Molecules (OMol25) dataset\add{,} released by FAIRChem\add{,} contains over 83 million unique molecules under 350 atoms \cite{levine_open_2025}, yet simply enumerating over just organic compounds up to 17 atoms yields 166 billion molecules, as demonstrated by the GDB-17 dataset \cite{ruddigkeit_enumeration_2012}. It is then an open question as to what \rep{the steps are}{are the steps} to obtain a model that learns patterns and laws in chemistry from just a fraction of possible molecules.

To address this issue, it has become routine to fine-tune uMLIPs. Fine-tuning is used in molecular modeling to incorporate system-specific data to uMLIPs while preserving the chemistry and physics captured while training, and introducing interactions relevant to the system of interest \cite{shenoy_role_2023, deng_systematic_2025, smith_approaching_2019}. This is done by fitting the models to new data, starting from a universal model fitted to a large dataset. Previous work has found that fine-tuning on just a few high-energy structures can eliminate this systematic error \cite{deng_systematic_2025}. However, this process is not standardized and varies across models \cite{liu_fine-tuning_2025}. For example, one approach to fine-tuning is multi-head fine-tuning \cite{kim_hydra_2023}, which preserves the main core (weights) of the model, while retraining only the final layers (heads). This approach is employed by the MACE family of models \cite{batatia_foundation_2024}, which we use in this paper. 

The first step to fine-tuning is to procure a dataset. One example of a fine-tuning data generation workflow to model liquid systems involves using a uMLIP to synthesize a dataset, labeling it with density functional theory (DFT), and fine-tuning on it to that data \cite{liu_fine-tuning_2025}. In this case, MD is used to augment data from a starting configuration and sample different bond lengths, different bond angles, and different local environments. More advanced sampling methodologies use techniques like uncertainty quantification\add{,}\cite{kulichenko_data_2024, hu_robust_2022} \add{ specialized data-selection techniques like DIRECT sampling, or} \cite{qiRobustTrainingMachine2024}\add{weighted-active space sampling }\cite{sealWeightedActiveSpace2025}\add{to maximize the coverage of the selected dataset. Enhanced sampling techniques such as umbrella sampling and on-the-fly probability enhanced sampling}\cite{yangReactantinducedDynamicsLithium2023} \rep{provide additional routes to maximize dataset coverage for synthetic datasets by biasing dynamics to capture rare and high-energy configurations that would otherwise be undersampled}{to maximize dataset coverage for synthetic datasets}. However, from \rep{the}{a} users' perspective, these techniques may be challenging \add{or nonobvious} to implement. To this end, relatively little work has been done to understand how these systematic underpredictions (biases) impact both the downstream liquid-state simulations and \add{the} fine-tuning process. 

In this paper, we investigate the implications of the systematically biased data sampled by uMLIPs. Specifically, we fine-tune \verb|MACE-MP-0b| on an entirely new chemical space to model the dynamics of a solution of choline-chloride and citric acid with dissolved divalent cobalt and lithium ions \cite{peeters_solvometallurgical_2020, batatia_foundation_2024}. To evaluate the effects of systematically biased data, we use two fine-tuning methods on solvent systems, and find that fine-tuning a uMLIP just once on system-specific data allows it to explore new atomic configurations \rep{that}{which} are never explored by the pre-trained uMLIP, which we demonstrate as a tightening of a bond length distribution. Models that are subsequently trained on this new information are then more accurate across different configurations and maintain accurate MD simulations. Significantly, models trained on only data sampled by the uMLIP are significantly less accurate with respect to potential energy predictions across a wide spread of configurations, and fail at accurately modeling MD, resulting in \rep{fictitious}{fictictious} reactions, which we unequivocally identify as results of extrapolations. 

\section*{Results}

Our first goal is to determine how systematic softening from uMLIPs \cite{deng_systematic_2025} impacts the quality of the fine-tuning dataset and eventual downstream fine-tuned model. To do this, we perform fine-tuning on a solution of choline chloride and citric acid, with dissolved divalent cobalt and lithium ions \cite{peeters_solvometallurgical_2020}. \rep{Neither component}{Both components} of the solvent \rep{appears}{do not appear} in the MPTrj dataset \cite{deng_chgnet_2023}, which is composed of solid crystalline materials, and is what \verb|MACE-MP-0b| was trained on. Thus, we are forcing the model to extrapolate new behaviors outside of its training dataset. With this realization, we use two methods to isolate the systematic softening effect on uMLIPs for liquid simulation.

\begin{figure}[H]
    \centering
    \includegraphics[width=1\linewidth]{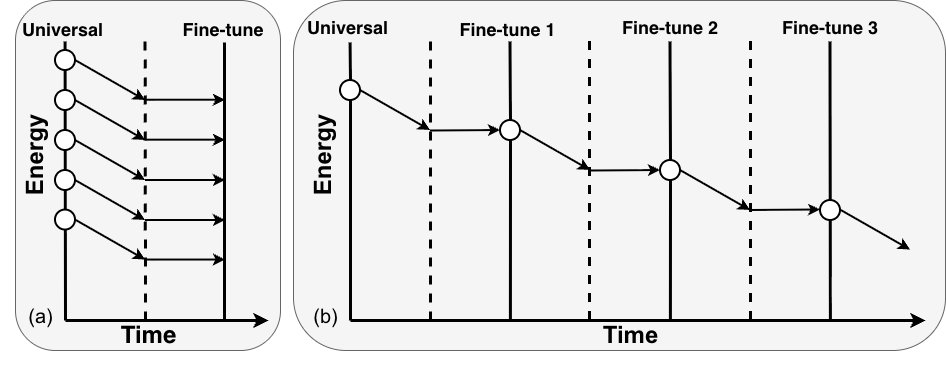}
    \caption{Schematic of the fine-tuning dataset generation strategies (energy not to scale). Each circle denotes a starting configuration for the most recent fine-tuned potential, \add{and} arrows represent MD trajectories, with equilibration and production runs separated by dashed lines. (a) The left panel describes the naive data generation workflow, where several independent trajectories are sampled in parallel to generate a dataset. (b) The right panel describes the \rep{iterative}{periodic} training workflow, where a single trajectory is sampled\add{,} and the potential is fine-tuned at fixed\del{-}intervals.}
    \label{fig:results_approaches}
\end{figure}

We first generate datasets for fine-tuning, with one dataset generated using only the uMLIP, which we call the naive dataset. The other fine-tuning dataset is generated by using a continuously fine-tuned model to accumulate through iterations, which we denote as the \rep{iterative}{periodic} dataset. Thus, we apply two workflows for data generation: A naive strategy\del{,} and an \rep{iterative}{periodic} strategy. A summary of each workflow can be seen in Figure~\ref{fig:results_approaches}. The left panel summarizes the naive fine-tuning workflow, where the five circles represent starting configurations \rep{that}{which} are used to initiate five MD trajectories using the uMLIP. This sampled data is accumulated into one dataset to fine-tune a single model. The right panel summarizes the \rep{iterative}{periodic} fine-tuning workflow, where a single circle denotes the initial configuration used to sample a single MD trajectory. Snapshots of the sampled trajectory are subsequently used to fine-tune the next model, and the process is repeated. \add{We choose to use a single trajectory to capture errors accumulating over longer timescales, and then use those datapoints to explicitly correct model errors in the next fine-tuning iteration. This sequential process emphasizes the difference in the quality of data sampled through fine-tuned models versus through a universal model. While another iterative workflow could restart from independent snapshots after each fine-tuning iteration, it may come at a cost of capturing more noise from Packmol generation of random structures.} In summary, we have two workflows that fine-tune models on the same amount of data, up to 50 points, generated from the same amount of MD simulation time, which is a total of 5 nanoseconds (ns). The two workflows differ only by their data generation procedure, where the naive workflow fine-tunes on data from a universal model, and the \rep{iterative}{periodic} workflow fine-tunes on data generated from iteratively fine-tuned models. 

To generate the datasets for fine-tuning, we start by initializing configurations using Packmol \cite{martinez_packmol_2009}. Next, we sample 1 ns NPT MD simulations at 300K and 0 bar, starting from each configuration, composed of 0.5 ns of equilibration, and 0.5 ns of production. We then generate labels as potential energy, atomic forces, and stresses from DFT calculations using both PBE and PBE+U (full computational details are in Methods). 

For naive fine-tuning, shown in Figure~\ref{fig:results_approaches}a, we initialize five starting configurations, and sample using MD for 1 ns (0.5 ns equilibration, 0.5 ns production) for each configuration. We denote naively fine-tuned models as N-Xpts, where X denotes the number of data points used to fine-tune the model. The values of X are chosen to match the number of data points to the \rep{iterative}{periodic} fine-tuning workflow, yielding the following models: N-10pts, N-21pts, N-31pts, N-40pts, N-50pts. 

For \rep{iterative}{periodic} fine-tuning, shown in Figure~\ref{fig:results_approaches}b, we initialize only one starting configuration and sample using MD for 1 ns (0.5 ns equilibration, 0.5 ns production), which is the same setup as for naive fine-tuning. However, we then fine-tune the universal potential on this data to generate FT1. For the next run, we take the last frame from FT1 as the starting configuration and sample using MD for 1 ns again (0.5 ns equilibration, 0.5 ns production), guided now by FT1. We repeat this process four more times, yielding models FT1 through FT5. The training dataset is cumulative across iterations; for example, the dataset used to fine-tune FT3 includes data generated with the universal potential, FT1, and FT2. The resulting models are denoted FT1, FT2, FT3, FT4, and FT5. \add{Thus, we directly compare the workflows by the number of fine-tuning configurations, e.g., FT1 is compared to N-10pts, FT2 is compared to N-21pts, etc.}

To test each model's performance on a variety of configurations, we independently generate a test set. We first \rep{generate}{randomly sample} structures using Packmol again and use \verb|MACE-MP-0b| to minimize the structures. Table~\ref{table:results_indTestSet} displays the model metrics (equations as root-mean-squared errors (RMSEs\add{, computed by equations}~\ref{eq:rmse_energy},~\ref{eq:rmse_force}\add{, and}~\ref{eq:rmse_stress}) of potential energies, forces, and stresses \add{and relative force errors (normalized by the mean absolute value of forces, computed by equation}~\ref{eq:relerror}\add{)}, when each model is evaluated on 22 independent test structures.
\begin{table}[H]
    \centering
    \caption{Model Metrics on Independent Test Set, N=22}
\begin{tabular}{lrrrrrrr}
    \hline
        \textbf{Metric} &RMSE\textsubscript{E} &RMSE\textsubscript{$F_x$} &RMSE\textsubscript{$F_y$} &RMSE\textsubscript{$F_z$} & \add{Rel. F Err} &RMSE\textsubscript{S}\\
        & meV/at & eV/\AA & eV/\AA & eV/\AA & \add{\%} & eV/\AA\textsuperscript{3} \\
    \hline
    \multicolumn{6}{l}{\textbf{Universal Model}} \\
        U & 21.24 & 0.223 & 0.221 & 0.229 & \add{175} & 0.0019  \\
    \hline
    \multicolumn{6}{l}{\textbf{N-X Models}} \\
        N-10pts & 10.30 & 0.201 & 0.206 & 0.214 & \add{161} & 0.0022 \\
        N-21pts & 11.34 & 0.193 & 0.196 & 0.204 & \add{155} & 0.0014 \\
        N-31pts & 11.37 & 0.191 & 0.192 & 0.202 & \add{152} & 0.0014 \\
        N-40pts & 9.02  & 0.187 & 0.188 & \textbf{0.197} & \add{149} & \textbf{0.0012} \\
        N-50pts & 10.19 & 0.186 & 0.187 & \textbf{0.197} & \add{\textbf{148}} & 0.0013 \\
    \hline
    \multicolumn{6}{l}{\textbf{FTX Models}}  \\
        FT1 & 10.06 & 0.206 & 0.209 & 0.216 & \add{160} & 0.0027 \\
        FT2 & 5.99  & 0.194 & 0.195 & 0.206 & \add{155} & 0.0026 \\
        FT3 & 6.78  & 0.189 & 0.190 & 0.201 & \add{151} & 0.0017 \\
        FT4 & 6.09  & 0.187 & 0.187 & 0.200 & \add{150} & 0.0019 \\
        FT5 & \textbf{5.79}  & \textbf{0.184} & \textbf{0.185} & 0.198 & \add{\textbf{148}} & 0.0017 \\
    \hline
\end{tabular}
\label{table:results_indTestSet}
\end{table}
Table~\ref{table:results_indTestSet} \rep{displays the aggregate RMSE metrics and relative force error for each fine-tuned model on an independent test set. We first note that the errors normalized to the mean absolute value of the forces in Table 1 appear large because the mean force in the independent test set is small (due to the minimization step).}{shows} The best model for predicting energy is FT5, which has an RMSE\textsubscript{E} of 5.79 meV/at. However, the \rep{iterative}{periodic} models, FT2-FT5, display similar accuracies at energy predictions. Notably, we observe a different trend for each workflow. For the \rep{iterative}{periodic} models, there is a significant reduction in RMSE\textsubscript{E} between FT1 and FT2, from 10.06 meV/at to 5.99 meV/at, which then plateaus. On the contrary, for the naive models, we observe that RMSE\textsubscript{E} has no correlation with the number of frames, and that the naive models perform significantly worse than their \rep{iterative}{periodic} counterparts, all averaging an RMSE of about 10 meV/at. We also observe that the total force errors are similar for the two workflows, and depend primarily on the number of training configurations, with larger datasets yielding lower errors. This behavior is expected since each configuration yields 3N force components (in this case, 50 frames yield\del{s} 29,550 force components), effectively weighting the training on the number of frames rather than their configurational diversity \cite{leimeroth_machine-learning_2025}. \add{We further decompose the relative force erorrs }

\begin{figure}[H]
    \centering
    \includegraphics[width=1\linewidth]{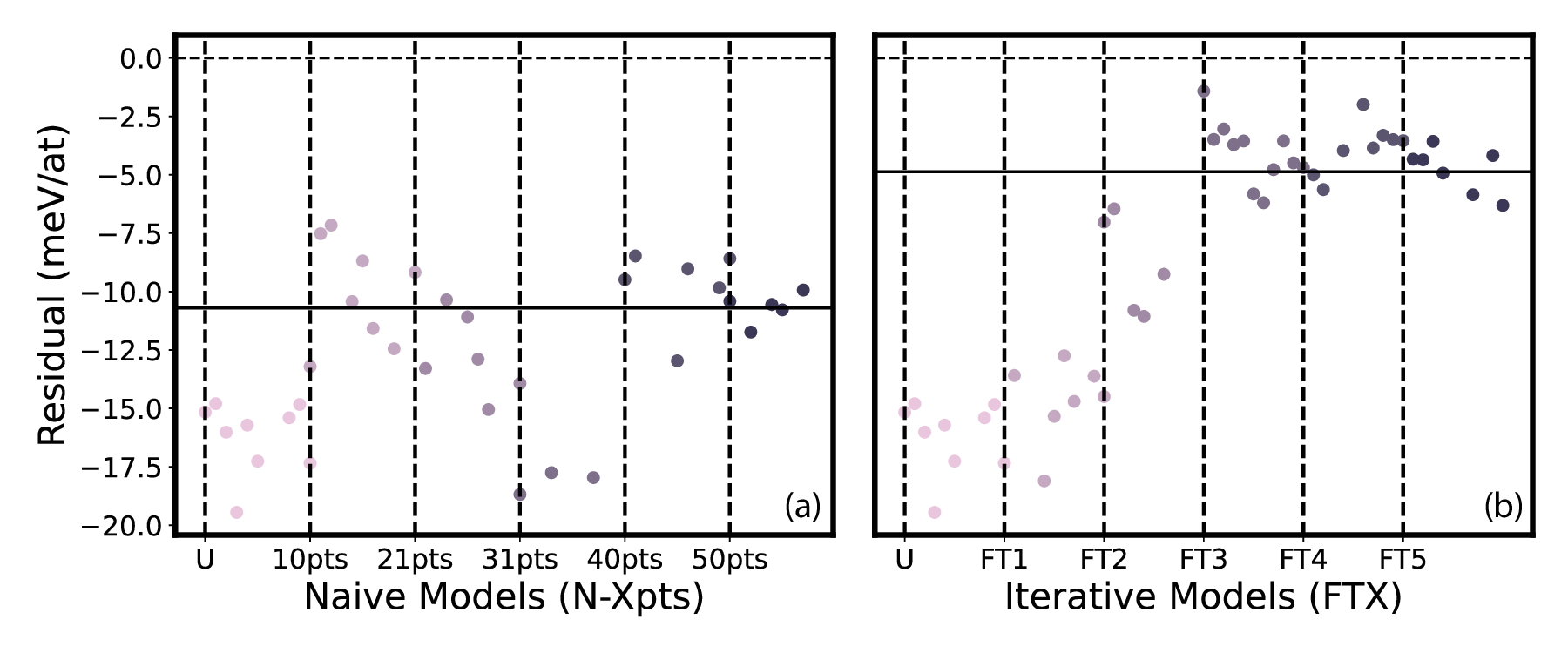}
    \caption{Testing errors during 0.5 ns of dynamics as residuals against DFT. The y-axis represents the residual as MD against DFT  (E\textsuperscript{MLIP}-E\textsuperscript{DFT}). For either workflow, the x-axis represents the model used to generate the trajectory, with colors to indicate different models. Lighter colors are models trained on fewer images, and darker colors are models trained on more images. The naive workflow (a) plateaus at about 10 meV/at, and the \rep{iterative}{periodic} workflow (b) plateaus at about -5 meV/at, as indicated by the horizontal black lines on each plot.}
    \label{fig:results_modelEvaluations}
\end{figure}

We next evaluate each model's performance during dynamics by allowing each model to sample a 1 ns trajectory starting from the same configuration. Figure~\ref{fig:results_modelEvaluations} shows both workflows' performance during MD as a potential energy residual (E\textsuperscript{MLIP}$-$E\textsuperscript{DFT}) against DFT. For each model, we sample 1 ns of MD (0.5 ns equilibration, 0.5 ns production). Configurations are collected every 50 ps, yielding 11 points including both endpoints. The snapshots are plotted in chronological order, with each color representing the results of a corresponding model during a 0.5 ns production run.  Each simulation (color) reflects a time evolution of the residual during MD.
Figure~\ref{fig:results_modelEvaluations} shows the performance of each workflow during dynamics, separated by model iterations. The naive workflow (Figure~\ref{fig:results_modelEvaluations}a) appears to give only modest improvements over the universal model, which has an average residual of about -16 meV/at, with fine-tuned models scoring average residuals of about -11 meV/at. Notably, it appears that N-31pts scores less than the uMLIP, scoring residuals of -18 meV/at. The \rep{iterative}{periodic} workflow (Figure~\ref{fig:results_modelEvaluations}b) shows continuous improvement up to FT3, which then plateaus at residuals of about -5 meV/at. Either workflow has a distinct effect on the downstream MD simulation, which will be discussed later. In either case, the \rep{iterative}{periodic} models are clearly more accurate than the naive models, as FT2 is more accurate than N-50pts. 

Figure~\ref{fig:results_modelEvaluations} is re-summarized in Table~\ref{table:results_evalModelMetrics}, along with the RMSEs in force and stress predictions on either model's self-generated test data.
\begin{table}[H]
    \centering
    \caption{Model Metrics on Self-Generated MD Test Set}
    \begin{tabular}{lrrrrrrr}
    \hline
    \textbf{Metric} &RMSE\textsubscript{E} &RMSE\textsubscript{$F_x$} &RMSE\textsubscript{$F_y$} &RMSE\textsubscript{$F_z$} & \add{Rel. F Err} &RMSE\textsubscript{S} & N \\
    & meV/at & eV/\AA & eV/\AA & eV/\AA & \add{\%} & eV/\AA\textsuperscript{3} & \\
    \hline
    \multicolumn{7}{l}{\textbf{Universal Model}} \\
    U & 16.28 & 0.175 & 0.157 & 0.182  & \add{26.8} & 0.0024 & 9\\
    \hline
    \multicolumn{7}{l}{\textbf{N-X Models}} \\
    N-10pts  & 10.24 & 0.154 & 0.153 & 0.158 & \add{24.7} & 0.0010 & 8 \\
    N-21pts  & 12.87 & 0.121 & 0.133 & 0.124 & \add{20.6} & 0.0013 & 6 \\
    N-31pts  & 18.14 & 0.163 & 0.152 & 0.133 & \add{23.5} & \textbf{0.0009} & 3 \\
    N-40pts  & 9.85  & 0.138 & \textbf{0.102} & 0.119 & \add{\textbf{18.7}} & 0.0010 & 6 \\
    N-50pts  & 10.70 & \textbf{0.115} & 0.144 & \textbf{0.116} & \add{19.5}& 0.0014 & 5 \\
    \hline
    \multicolumn{7}{l}{\textbf{FTX Models}} \\
    FT1      & 14.74 & 0.155 & 0.149 & 0.160 & \add{24.4}& 0.0013 & 7 \\
    FT2      & 9.19  & 0.136 & 0.137 & 0.127 & \add{21.2}& 0.0013 & 5 \\
    FT3      & 4.26  & 0.137 & 0.134 & 0.134 & \add{22.0}& 0.0013 & 11 \\
    FT4      & \textbf{3.98}  & 0.130 & 0.134 & 0.134 & \add{21.4} & \textbf{0.0009} & 8 \\
    FT5      & 4.87  & 0.125 & 0.129 & 0.125 & \add{19.6} & 0.0016 & 7 \\
    \hline
    \end{tabular}
    \label{table:results_evalModelMetrics}
\end{table}

\add{Table}~\ref{table:results_evalModelMetrics}\add{ shows the metrics of each model at each fine-tuning step, where N is the number of converged DFT evaluations from the sampled MD trajectory}. Table~\ref{table:results_evalModelMetrics} again shows that energy RMSEs do not appear to change with the number of data points in the naive approach\add{,} as all RMSE values are closely distributed around 10 meV/at, \rep{except for}{with the exception of} N-31pts, which is an outlier with an RMSE of 18.14 meV/at. This may be due to the number of points (N=3), which is low because not all sampled MD frames converged. On the other hand, we observe a steady increase in accuracy from FT1 to FT4, where RMSE decreases from 16.28 meV/at to 3.98 meV/at. As was seen in the independently generated test set metrics in Table 1, force errors appear to depend on the number of frames rather than the workflow, as both cases have a trend of RMSE\textsubscript{F} decreasing from around 0.15~eV/A to around 0.12~eV/at as more points are added.

To understand the apparent differences in the learning process between the naive and \rep{iterative}{periodic} models, we next decompose each model's dataset into smooth-overlap-of-atomic-positions (SOAP) descriptors and perform a principal component analysis (PCA) to observe how the datasets compare in chemical space. One PCA is fit to the concatenated datasets, totaling 100 points, and the first two 2 principal components (PCs) capture 41.82\% of explained variance, with five PCs capturing 68.91\% of the explained variance.

\begin{figure}[H]
    \centering
    \includegraphics[width=1\linewidth]{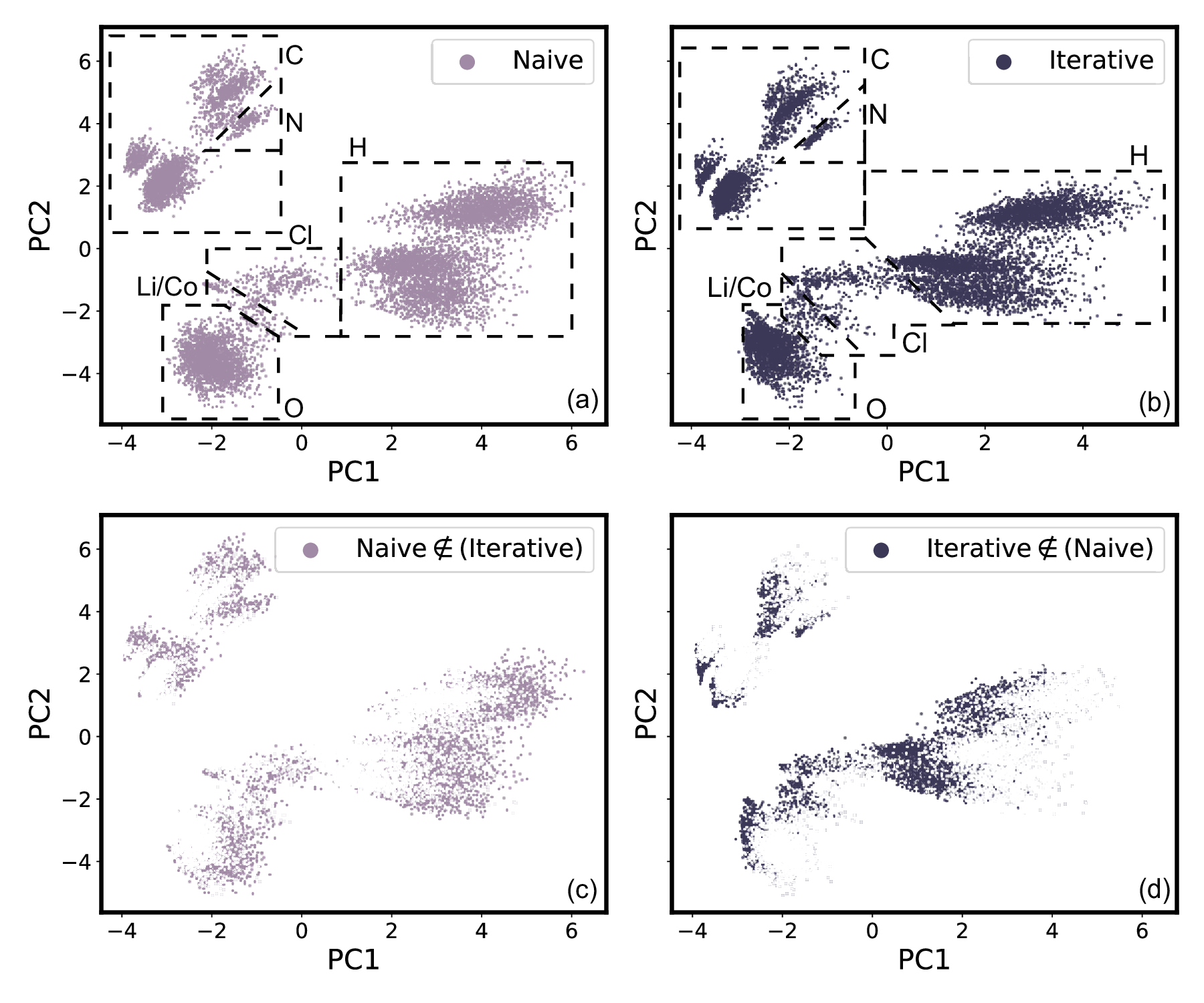}
    \caption{Principal component analysis of the individual 50-point datasets (naive or \rep{iterative}{periodic}), with regions labeled by atom type. The first two principal components account for 41.82\% of explained variance. Panels (a) and (b) show the full datasets, while panels (c) and (d) display only the unique regions sampled by each method, which is represented visually by overlaying white coloring on top of the corresponding dataset, constituting\add{,} for example, Naive$\notin$(\rep{Iterative}{Periodic}). The left panels (a, c) correspond to the naive workflow, and the right panels (b, d) correspond to the \rep{iterative}{periodic} workflow.}
    \label{fig:results_PCA_Datasets}
\end{figure}

Figure~\ref{fig:results_PCA_Datasets} depicts the PCA of either dataset's SOAP descriptors using the first two PCs, the regions where each atom type resides in PC space (labelled in dashed boxes), and the unique regions of either dataset relative to each other, denoted as Naive \( \notin \) \rep{Iterative}{Periodic} or \rep{Iterative}{Periodic} \( \notin \) Naive. 

Qualitatively, the clusters separate into elemental groups\del{,} and further into distinct chemical environments. Each dashed box denotes a distinct atom type, with subclusters within each box corresponding to different chemical environments.  Distinctly, there are three larger clusters of data, which correspond to carbon, hydrogen\add{,} and oxygen environments. These occupy a majority of the data. Within each element cluster, there are subclusters that map to different chemical environments. In the top left of the plot, for example, carbon atoms form four distinct subclusters corresponding to the \ce{N-CH3}, \ce{R-COOH}, \ce{R-COH}, and \ce{R-CH3} environments. Next to this\del{,} is the nitrogen environment, which is a single cluster, consistent with nitrogen appearing only in choline and being bonded to four carbon atoms. The oxygen environments represent one homogenous cluster towards the bottom left of the plot, indicating that these environments are similar, as described by SOAP descriptors, throughout the dataset. The hydrogen cluster represents a majority of the data\del{,} and resembles the right-hand side of the plots. Here, since hydrogen only bonds to single atoms, the different subclusters represent the \ce{H-O}, \ce{H-C}, and, to a much lesser extent, \ce{H-Cl} bonds.  

While both datasets have comparable clusterings, as both datasets share the same composition and environments, they differ in the regions and shapes, as indicated by the bottom panels of Figure~\ref{fig:results_PCA_Datasets}. These regions resemble the differences in the composition of local environments in each dataset. We observe that the naive sampling covers a more diverse set of data through random generation, resulting in diffuse, sparse regions. More discussion on this sampling will follow. Additionally, the \rep{iterative}{periodic} dataset appears to span larger regions along PC1.

\begin{figure}[H]
    \centering        
    \includegraphics[width=1\linewidth]{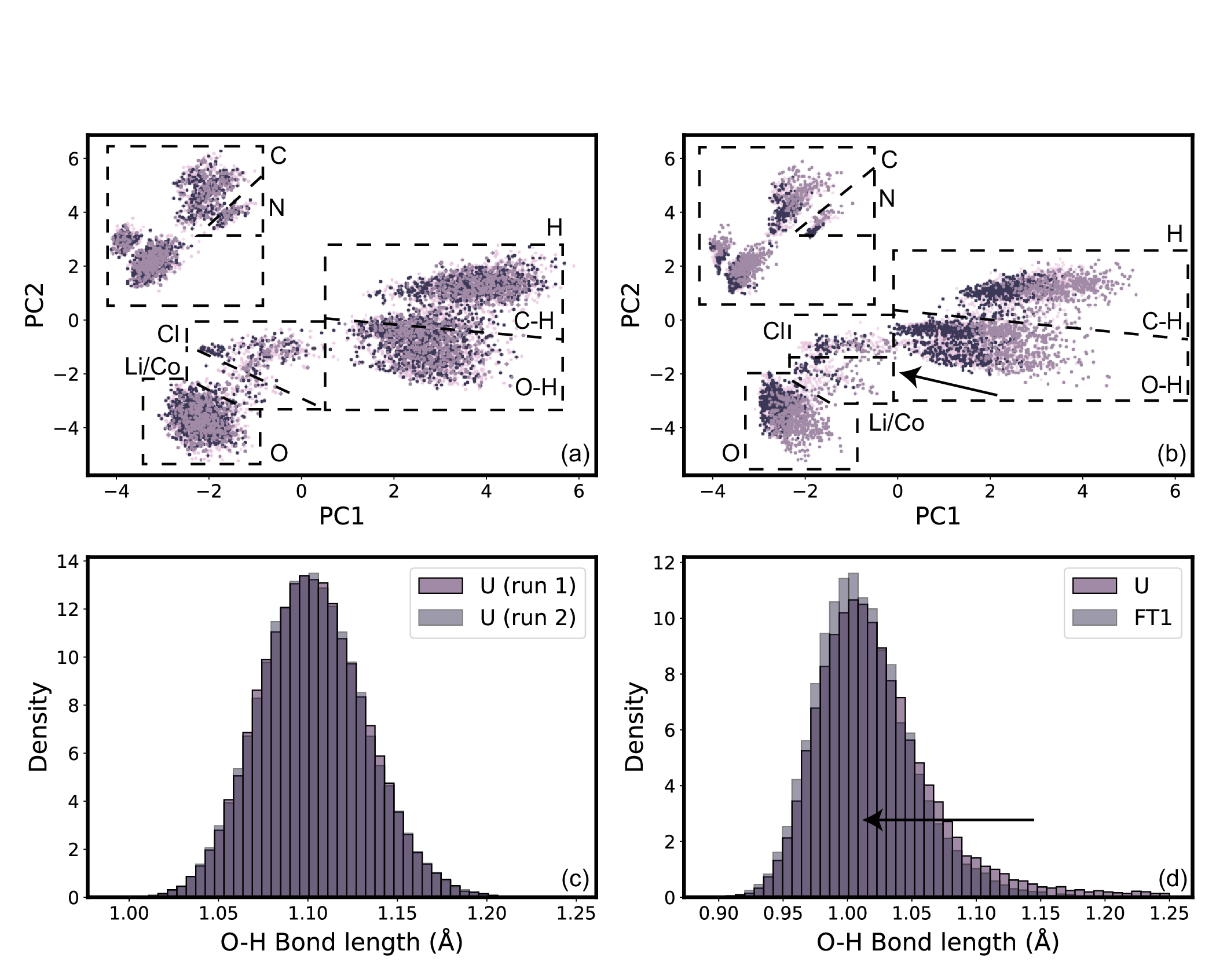}        
    \caption{Principal component analysis of each dataset from the naive and \rep{iterative}{periodic} approaches, with regions labeled by atom type, and bond-length distributions of O-H bonds. \rep{Light purple represents data sampled from the universal potential, and dark purple represents data sampled from N-10pts and FT1, and pink represents data sampled by later iterations}{Each color represents the unique coverage contributed by the most current iteration (lighter colors are earlier models)}. Panel (a) corresponds to datasets used to \rep{fine-tune}{generate} N-10pts, N-21pts, N-31pts, N-40pts, and N-50pts\add{,} respectively. Panel (b) likewise corresponds to datasets used to \rep{fine-tune}{generate} FT1, FT2, FT3, FT4, and FT5. The arrows in panel (b) represent movement from data sampled from the universal potential to data sampled from FT1 to FT5. Panels (c) and (d) represent the \add{two} distributions of O-H bond lengths between \rep{two independent trajectories sampled}{data generated} by (c) only the universal model\add{,} and (d) the universal and a fine-tuned model. Panel (d) relates the bond length distribution to the PCA distribution shift with arrows.}
    \label{fig:results_PCAPerIter}
\end{figure}

Next, each dataset is visualized in terms of how PC space is traversed between each iteration. Figure~\ref{fig:results_PCAPerIter} illustrates how each workflow explores the configuration space. Each iteration is decomposed into its unique coverage, as compared to previous iterations. Lighter colors represent earlier datasets, and are plotted on top. Darker colors represent later datasets and are plotted \rep{at the}{on} bottom. Thus, each panel represents the dataset growth per iteration, and each point represents the unique growth covered by that dataset. The left plot represents the naive workflow, and the right plot represents the \rep{iterative}{periodic} workflow. We also include distributions of \ce{O-H} bond length distributions to demonstrate the difference between the two sampling methods, which contributes to the differences we observe.
Figure~\ref{fig:results_PCAPerIter} shows the shift of each dataset in PC space over each iteration of the fine-tuning process, with histograms to demonstrate differences in bond lengths through different iterations. The naive workflow (Figure~\ref{fig:results_PCAPerIter}a) saturates almost immediately, characterized by the exploration occurring primarily through expanding the covariance ellipse, or outwardly growing elliptical clouds. In contrast, the \rep{iterative}{periodic} workflow (Figure~\ref{fig:results_PCAPerIter}b) first shifts from the initial training configuration, as indicated by the arrow from the pink dataset to newer datasets, followed by a saturation indicated by dense regions of dark colors. This shift represents a combination of several factors that are captured in SOAP parameters, including local environments, bond lengths, and bond angles. We demonstrate this in Figure~\ref{fig:results_PCAPerIter}c and ~\ref{fig:results_PCAPerIter}d, which illustrate the distributions of \ce{O-H} bond lengths. Here, the distribution tightens as shorter \ce{O-H} bonds form, as shown in Figure~\ref{fig:results_PCAPerIter}c and ~\ref{fig:results_PCAPerIter}d. We also notice this \rep{phenomenon}{phenomena} in \ce{H-Cl} and \ce{C-O} bonds, where universal potentials generate broad distributions that are softened around the peaks (Figure ~\ref{fig:SI_C-O_BondLengths}, ~\ref{fig:SI_H-Cl_BondLengths}). These results indicate that there is a systematic difference between the configurations explored by fine-tuned models\del{,} and the universal potential.

\begin{figure}[H]
    \centering
    \includegraphics[width=1\linewidth]{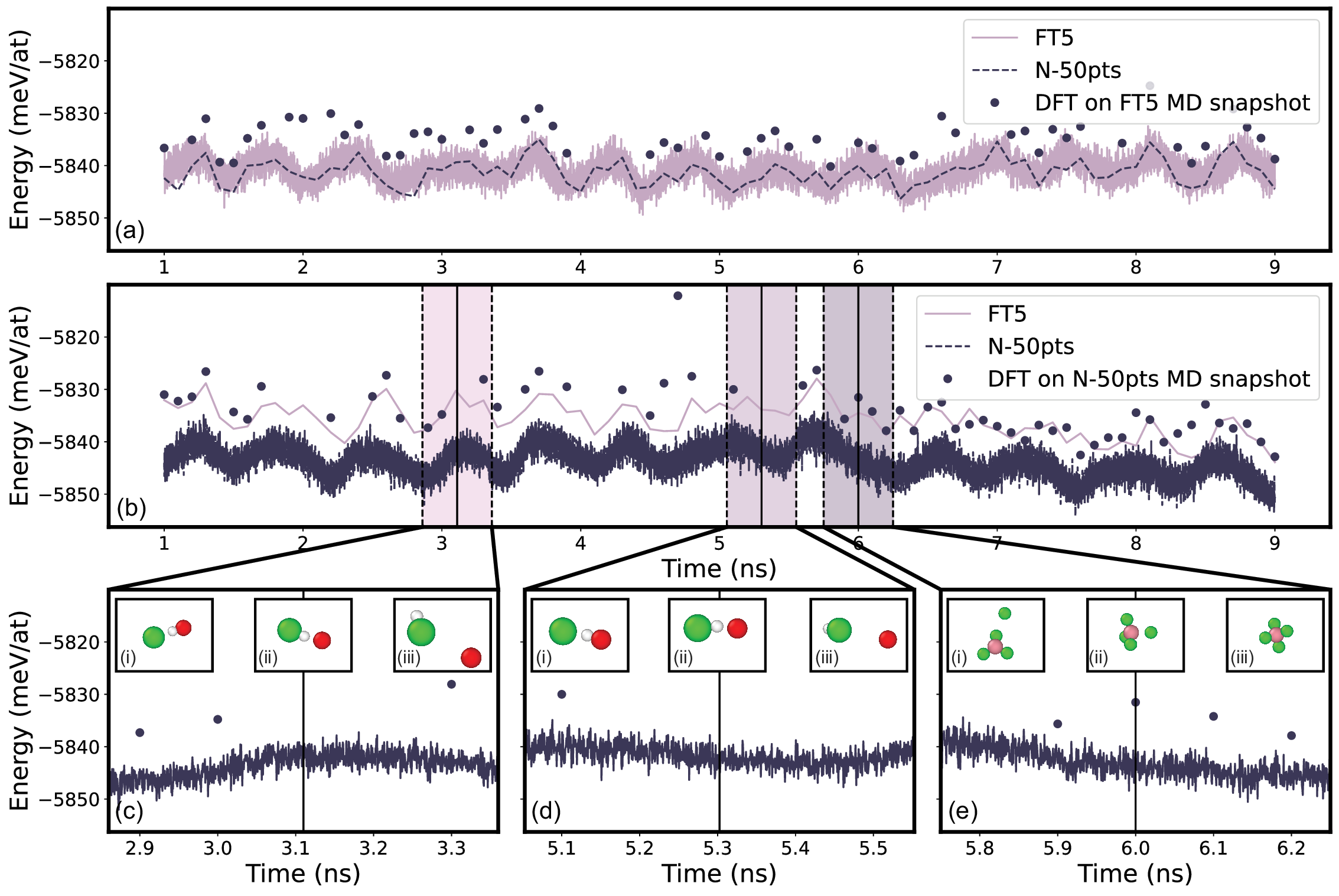}
    \caption{A summary of 8 ns production trajectories \rep{sampled}{generated} by N-50pts and FT5. Panels (a) and (b) summarize the 8 ns production trajectories in terms of potential energy, with model\add{-}predicted energy (solid line), cross-evaluated energy (dashed line), and \rep{single-point DFT evaluations of trajectory snapshots}{DFT references} (circles) sampled every 0.1 ns for FT5 (a) and N-50pts (b), respectively. Panels (c, d, e) highlight time ranges t\textsubscript{1}=[2.86, 3.36] ns,  t\textsubscript{2}=[5.0525, 5.5525], and t\textsubscript{3}=[5.75, 6.25], during the sampling of N-50ps (b), where artifacts occur. For each range, representative snapshots illustrate three figures of the artifacts: panels (c-d) illustrate deprotonation reactions between chlorine (green), oxygen (red), and hydrogen (white), while panel (e) demonstrates a change in cobalt's (pink) coordination environment from \ce{CoCl3} to \ce{CoCl4}.}
    \label{fig:results_longEval}
\end{figure}

As a final test of dynamics, we let the final models from each workflow, N-50pts and FT5, sample a trajectory for 9 ns (1 ns \rep{equilibration}{equilibriation}, 8 ns production) starting from the same Packmol-generated configuration.  Figure~\ref{fig:results_longEval}a and ~\ref{fig:results_longEval}b show the potential energy trends over the production run (1 to 9 ns) for each trajectory as model predictions, cross-evaluations, and the DFT references. During the naive trajectory, three reactions are observed at $t_1=3.16$ ns, $t_2=5.3$ ns, and $t_3=6$ ns, and these instances are illustrated in Figure~\ref{fig:results_longEval}c, ~\ref{fig:results_longEval}d, and ~\ref{fig:results_longEval}e, where we plot configurations before, during, and after the reactions. They correspond to artifacts such as deprotonation reactions and changes \add{in coordination} environment \del{changes} for cobalt.  
Figure~\ref{fig:results_longEval} highlights the differences observed between N-50pts and FT5 in a long (8 ns) simulation. Figure~\ref{fig:results_longEval}\rep{b}{a} shows the accuracy of N-50pts against DFT. First, we find several deprotonation reactions occur, which correspond to the highlighted regions $t_1$ and $t_2$ in Figure~\ref{fig:results_longEval}\rep{b}{a}, and illustrated in Figure~\ref{fig:results_longEval}c and ~\ref{fig:results_longEval}d. In both of the deprotonation reactions, which occur between different hydrogens and chlorines, chlorine bonds to an oxygen. In addition, we observe two regimes, where N-50pts is significantly less accurate before 6 ns, evaluating at an RMSE\textsubscript{E} of 15 meV/at before, and 10 meV/at after. After about 6 ns, the reaction in \ref{fig:results_longEval}e occurs, where \ce{CoCl3} bonds to \ce{Cl}, forming \ce{CoCl4}, which is a known silvation environment for Co(II). In contrast, \add{Figure}~\ref{fig:results_longEval}\add{a reveals that} FT5 consistently samples at RMSEs of about 5 meV/at throughout the trajectory. We also perform a cross-evaluation, where we allow each model to evaluate the trajectory generated by the other model. Table~\ref{table:results_crossEvals} summarizes the potential energy, force, and stress prediction RMSEs for N-50pts, FT5, and their cross evaluations.
\begin{table}[H]
    \centering
    \caption{FT5 and N-50pts 9 ns trajectory metrics}
    \begin{tabular}{lrrrrrrr}
        \hline
        \textbf{Metric} & RMSE\textsubscript{E} & RMSE\textsubscript{$F_x$} & RMSE\textsubscript{$F_y$} & RMSE\textsubscript{$F_z$} & \add{Rel. F Err} & RMSE\textsubscript{S} & N \\
        & meV/at & eV/\AA & eV/\AA & eV/\AA & \add{\%} &   eV/\AA\textsuperscript{3} & \\
        \hline
        \textbf{Regular Eval.} \\
        FT5      & \textbf{6.41}  & 0.121 & \textbf{0.109} & 0.124 & \add{19.0} & \textbf{0.0011} & 56 \\
        N-50pts  & 10.48 & \textbf{0.105} & 0.119 & \textbf{0.114} & \add{\textbf{18.3}} & 0.0014 & 58 \\
        \hline
        \textbf{Cross Eval.} \\
        FT5 on N-50pts & \textbf{4.16} & 0.109 & 0.114 & 0.113 & \add{18.1} & 0.0016 & 58 \\
        N-50pts on FT5 & 6.78 & \textbf{0.097} & \textbf{0.101} & \textbf{0.102} & \add{\textbf{16.1}} & \textbf{0.0011} & 56 \\
        \hline
    \end{tabular}
    \label{table:results_crossEvals}
\end{table}
Presently, we first observe that FT5 achieves more accurate energy predictions on both trajectories, with no apparent difference between force and stress predictions, which evaluate at roughly 0.1 eV/\AA, and 0.001 eV/\AA\textsuperscript{3} respectively, and follow from the above observations. These findings are consistent with Table~\ref{table:results_evalModelMetrics}, as N-X models typically perform at energy errors of 10 meV/at on MD structures, and FTX models perform at around 5 meV/at on MD structures. \add{Furthermore, we decompose forces by element type in Table}~\ref{tab:SI_per_element_comparison}\add{, where we observe errors of about 13-18\% (~0.06 eV/Å-0.13 eV/Å) in the bulk solvent (C, N, O, H), which include about 96\% of the system, with larger errors of approximately 35-100\% in the solvated ions (Cl, Li, Co). We include percentile statistics in Table}~\ref{tab:SI_per_element_distribution}\add{, where we observe that N-50pts displays higher error variance for the bulk solvent.}

Additionally, FT5 samples a trajectory \rep{without}{that does not contain any model} artifacts because there are no broken bonds. In contrast, the N-50pts samples several deprotonation reactions, two of which are in Figure~\ref{fig:results_longEval}c and \ref{fig:results_longEval}d, and \rep{have}{has} inconsistent energy errors throughout the trajectory\add{,} which appear to align with a change in cobalt's coordination environment that occurs at 6 ns.  Energy predictions before 6 ns are considerably less accurate than those after 6 ns. Table \ref{table:prediction_regime_table} illustrates this difference, where we evaluate the accuracy of both FT5 and N-50pts on the long naive-generated trajectory before and after 6 ns have elapsed.
\begin{table}[H]
        \centering
        \caption{Energy Errors in Different Evaluation Regimes}
        \begin{tabular}{lrrr}
        \hline
        RMSE\textsubscript{E} & N-50pts & FT5 & N \\
        \hline
         & meV/at & meV/at &  \\
        \hline
        Before 6ns & 12.7 & \textbf{6.45} & 27 \\
        After 6ns    & 7.99 & \textbf{6.32}& 31 \\
        \hline
    \end{tabular}
    \label{table:prediction_regime_table}
\end{table}
Indeed\add{,} the N-50pts demonstrates model artifacting indicated by a decrease in RMSE from 12.70 to 7.99 meV/at, compared to a consistent error throughout the trajectory using FT5. This indicates that after 6 ns, the dynamics begin to sample a new domain that is within the dataset of N-50pts, moving from extrapolation (indicated by high, varying errors) to interpolation (indicated by low, consistent errors).

\begin{figure}[H]
    \centering
    \includegraphics[width=1\linewidth]{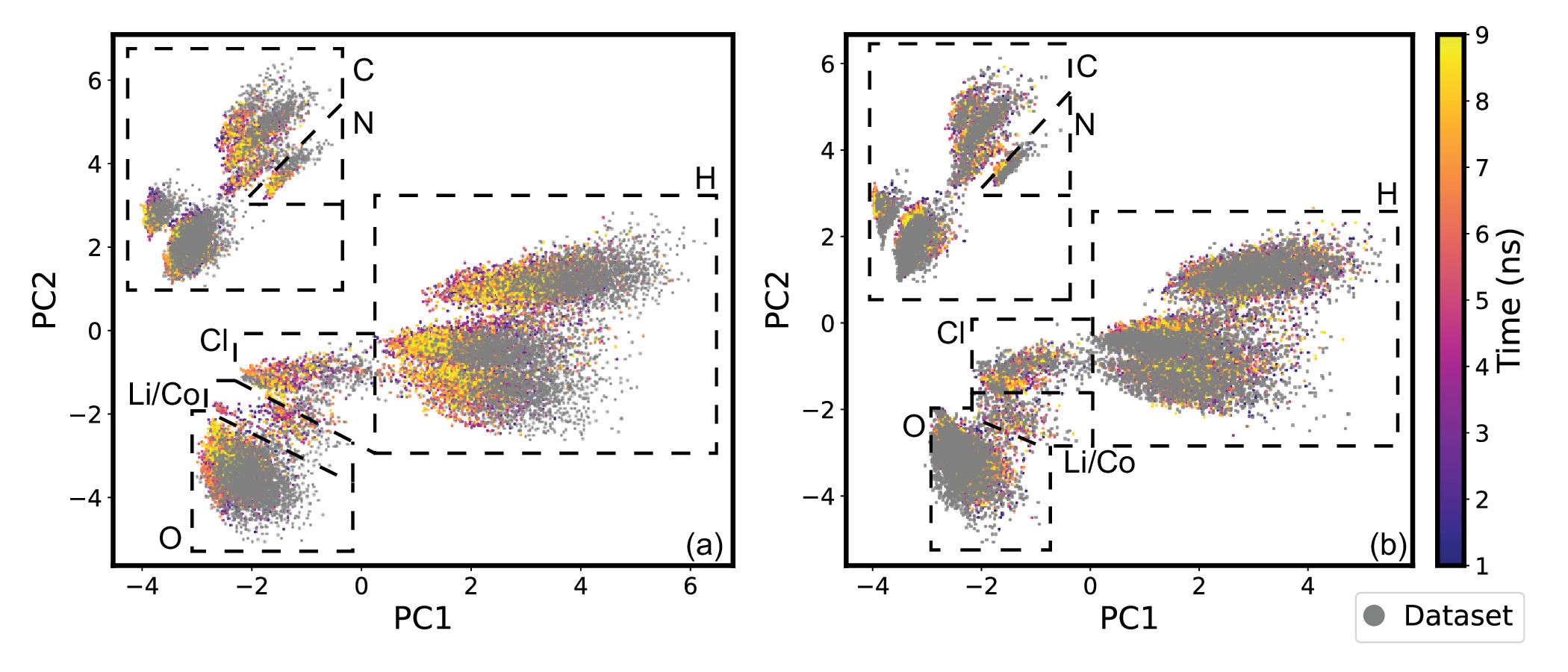}
    \caption{Time-resolved principal component analysis for the 9 ns trajectories sampled by (a) N-50pts and (b) FT5, with regions labeled by atom type. Each panel represents 8 ns of simulation, sampled at intervals of 0.1 ns, projected onto PC space, represented as colored points. Purple indicates earlier frames, and yellow represents later frames. For comparison, the respective dataset is also projected onto PC space (grey).}
    \label{fig:results_timeResolvedPCA}
\end{figure}

We next investigate potential sources for the artifacts sampled by N-50pts. First, we project all images from each model's sampled trajectory onto PC space and its space as related to each model's dataset. Figure ~\ref{fig:results_timeResolvedPCA} describes the progression of each trajectory with images taken at 0.1 ns intervals, and colors represent the progression of time from 1 to 9 ns. 
Figure~\ref{fig:results_timeResolvedPCA} shows the time-resolved PCA for each trajectory, and summarizes how each model samples MD. The naive model consistently samples outside of its dataset, indicated by the colored areas in panel~\ref{fig:results_timeResolvedPCA}a. Through the color gradient, specifically in the hydrogen (right\add{-}hand side) regions, we notice that the lighter regions are closer to the dataset (grey) than \add{the} darker regions, indicating that the trajectory moves closer to the dataset throughout the trajectory. This aligns with the observations previously observed in Table~\ref{table:prediction_regime_table}, where later evaluations end up more accurate. In contrast, the \rep{iterative}{periodic} model does not display any patterned errors, and consistently evaluates on configurations close to the dataset (grey). 

\add{The remainder of this study, presented in the Discussion and Supplementary Information, extends our findings to a third workflow and an aqueous system, providing further analyses that interpret the differences between iterative and naive fine-tuning.}
\section*{Discussion}

This study isolates uMLIP bias and its effects on the quality of machine-learned molecular dynamics. Specifically, we find that when presented with out-of-domain systems, MACE uMLIPs exhibit bias in molecular dynamics trajectories, limiting the quality of configurations collected for fine-tuning, resulting in a relatively limited accuracy of downstream fine-tuning tasks to about 10 meV/at. A potential solution we report is to fine-tune multiple times:  Figure~\ref{fig:results_PCAPerIter}b demonstrates that after fine-tuning on ten frames, the FT1 potential already begins to explore configurations unseen in the five uMLIP-sampled trajectories in the naive workflow, indicated by the arrows from the configurations in pink. This trend continues in subsequent models, where FT2, FT3, FT4, and FT5 \add{are} fine-tuned on a previous model's new trajectories demonstrate increased accuracy at predicting energies on both unknown configurations (Table~\ref{table:results_indTestSet}), and during MD (Table~\ref{table:results_evalModelMetrics}). We observe that configurations collected by the uMLIP through the naive workflow exhibit\del{s} a pattern where exploration occurs outward in PC space. By comparing these configurations with those spanned by simulations in (Figure~\ref{fig:results_PCAPerIter}a), we notice that the diffuse exploration is relatively unmeaningful, as the configurations do not sample overlapping distributions. For this reason, we conclude that uMLIPs do not sample representative configurations on new domains. Instead, they tend to bias MD toward certain configurations, which may be in the form of local environments, bond angles, or bond lengths, as evidenced by Figure~\ref{fig:results_PCAPerIter}c. Another consideration is \add{the} length of MD, which we consider by training on an extended 6.2 ns-long trajectory, (training and evaluation in SI, Figs. ~\ref{fig:SI_noniterative_workflow}, ~\ref{fig:SI_indEvalSet}, ~\ref{fig:SI_modelEvaluationsNoniterative}, ~\ref{fig:SI_PCA_DatasetsNoniterative}, and~\ref{fig:SI_results_PCAPerIterNoniterative}). We find that this procedure results in overfitting\del{,} and failure to generalize to a variety of configurations, which we discuss later in this section. A direct consequence of a fine-tuned uMLIP on uMLIP-sampled trajectories is that downstream models then produce unphysical dynamics. Figure~\ref{fig:results_longEval} describes these symptoms, which can be fictitious bond breaking/formation events such as deprotonation and HCl generation, long periods ($\sim$6 ns) of extrapolation reflected by higher energy RMSEs, and unphysical solvation shells, namely \ce{CoCl3} instead of \ce{CoCl4}.

While uMLIPs promise near-ab initio accuracy at a fraction of the cost\cite{unke_machine_2021} while also lowering the technical barrier to deploying quantum-mechanically accurate MD simulations \cite{leimeroth_machine-learning_2025, liu_fine-tuning_2025}, they extrapolate poorly on structures outside of their training manifold \cite{loewUniversalMachineLearning2025b, kreiman_understanding_2026}. It is known that out-of-domain evaluations manifest as a systematic softening, or systematic biases, of the PES, which can be partially mitigated through fine-tuning \cite{deng_systematic_2025}. Here, we identify this as a bias that arises from domain shift, which we will discuss next, as we apply uMLIPs to structures outside of their training manifold. 

We identify this bias in Figure~\ref{fig:results_PCAPerIter} in the form of a softening of the bond length distribution before fine-tuning. Because the uMLIP is prompted with domains it is not familiar with, it biases MD and local molecular structures. We observe that the distribution of \ce{O-H} bond lengths resemble\add{s} a wider distribution when sampled from the universal potential. We have also identified this \rep{phenomenon}{phenomena} for \ce{H-Cl} bonds, and \ce{C-H} bonds, where the uMLIP also yields a tighter distribution of bond lengths (Figure~\ref{fig:SI_C-O_BondLengths}, ~\ref{fig:SI_H-Cl_BondLengths}). This indicates that uMLIP-sampled structures are too flexible around equilibrium. Due to this, the majority of the difference between the two workflows occurs in the first two iterations, which is also apparent in Table~\ref{table:results_evalModelMetrics}. 

Thus, our study indicates additional considerations beyond training on high-energy configurations as recommended by Deng et al. for reliably fine-tuning uMLIPs \cite{deng_systematic_2025}. Specifically, the quality of the fine-tuning dataset influences downstream energy and force predictions. Fine-tuning on only data generated from uMLIPs generates noisier datasets that lead to extrapolations and unphysical MD simulations. Our work finds that generating a model for accurate liquid simulations requires an iterative workflow: fine-tuning a universal potential, running a trajectory with the fine-tuned model, and fine-tuning on the newly sampled configurations. This methodology ensures representative sampling and yields more accurate models than relying on uMLIP sampling alone.

We further investigate the mechanisms that underlie the artifacts in the 9 ns trajectory sampled by N-50pts, or the examples illustrated in Figure~\ref{fig:results_longEval}c, Figure~\ref{fig:results_longEval}d, and Figure~\ref{fig:results_longEval}e. We hypothesize that these artifacts are due to erratic force predictions in low force regimes, which are caused by extrapolation errors. Specifically, we analyze the forces on hydrogen atoms, which appear to be the source of the results, and surmise that erroneous predictions are due to extrapolation errors. 

\begin{figure}[H]
    \centering
    \includegraphics[width=1\linewidth]{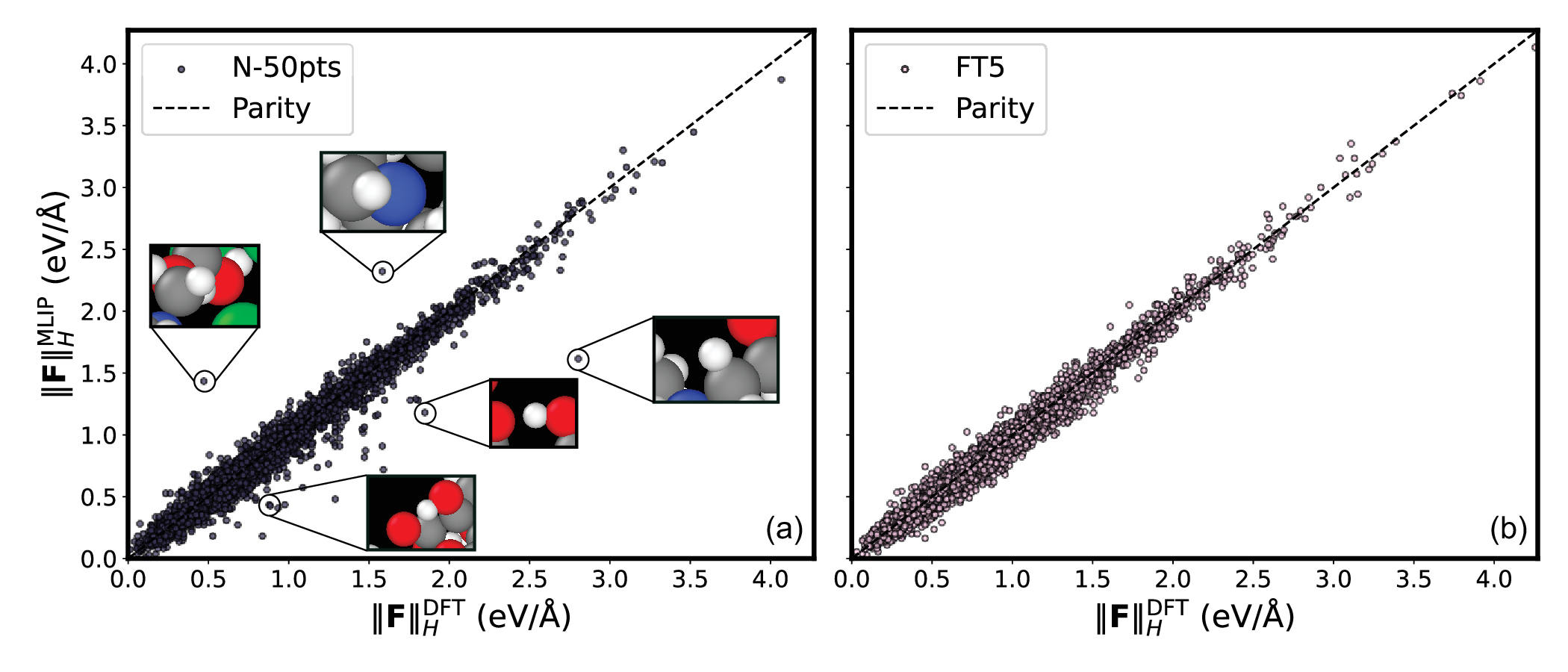}
    \caption{A parity plot between the magnitudes of force predictions on hydrogen atoms from (a) N-50pts (N=5220) and (b) FT5 (N=5040), and DFT. Results below the parity line represent MLIP force underpredictions, and points above the line represent force overpredictions. Panel (a) also contains inlays describing several hydrogen environments with poor force predictions. The inlays contain elements as circles, including hydrogen (white), carbon (gray), oxygen (red), chlorine (green), and nitrogen (blue).}
    \label{fig:discussion_H_forces}
\end{figure}

In the following sub-study, we observe the chemical environments and configurations in relation to the fine-tuning dataset. We aim to show that the structures captured by the universal potential are not sufficient to accurately model the system, and these reactions arise from this insufficiency. We first compare the predicted hydrogen forces during MD to DFT. Figure~\ref{fig:discussion_H_forces} is a parity plot that represents the quality of MLIP predictions compared to DFT during the 9 ns trajectory generated by both models of only hydrogen atoms.
Figure~\ref{fig:discussion_H_forces} illustrates the error in the magnitudes of force predictions between FT5, N-50pts, and DFT in the trajectory sampled by N-50pts, with select environments with poor predictions represented in inlays.  Figure~\ref{fig:discussion_H_forces}a represents predictions made by N-50pts, and Figure~\ref{fig:discussion_H_forces}b represents predictions made by FT5. We observe that N-50pts exhibits inaccuracies of up to 1.0 eV/\AA \ at force magnitudes between 0.0 and 3.0 eV/\AA \ whereas FT5 is consistently accurate throughout the simulation at all \rep{hydrogen force magnitudes}{forces}. Interestingly, a number of poor predictions occur for methyl-hydrogens within choline. High force errors also appear for \ce{O-H} groups, where \del{often} these forces are often underpredicted by the MLIP, as characterized by several of the inlays.

To relate the dataset to the trajectories, we use Q-residuals, or squared prediction errors, which are a statistical outlier metric. They have been used in image processing to find outlier pixels \cite{jablonski_principal_2015} and statistical process control for fault detection \cite{kourti_application_2005}. Q-residuals are a lack-of-fit metric for PCA, and measure the reconstruction error of any two given datasets. Specifically, we fit a PCA model to the SOAP descriptors of the naive dataset\del{,} and use this model to reconstruct descriptors from the trajectories. The Q-residual is defined as the distance between the reconstructed and original descriptors. Each atom in each frame is assigned a Q-residual that measures how well its local environment is represented by the dataset, where larger Q-residuals indicate environments that lie outside the training distribution. Thus, Q-residuals represent a computationally efficient measure of departure from \add{the} distribution.

\begin{figure}[H]
    \centering
    \includegraphics[width=1\linewidth]{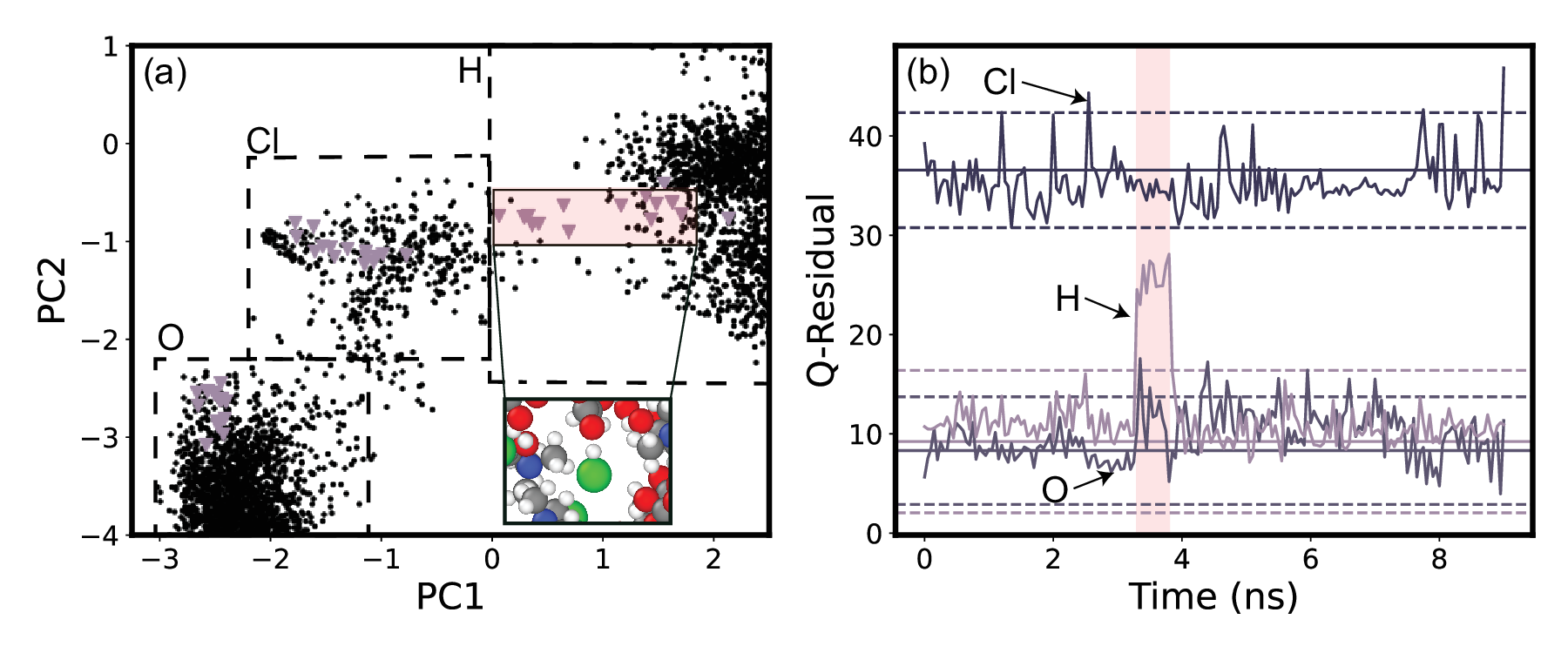}
    \caption{Example of Q-residuals indicating extrapolatory behavior during a 9-ns long N-50pts MD simulation between 3.3 and 3.8 ns. (a) Principal component analysis (PCA) on SOAP descriptors shows the variance of environments spanned by different elements, Cl, H, and O. Purple indicates environments sampled by the simulation from 3.3 to 3.8 ns, and black represents the environments spanned by the fine-tuning dataset.  The inset illustrates the deprotonation \del{that} of a H (white circle) from an O (red) \del{occurs,} forming an HCl. (b) Q-residuals on the specific H, Cl, and O atoms \rep{quantify}{quantifies} the moment where extrapolation occurs (red box), where the deprotonating H bonds with Cl to form HCl. We plot the mean of each atom-type as a solid line of the corresponding color, with dashed lines indicating $\pm$ 2 standard deviations.}
    \label{fig:discussion_rxn_3ns_qresidual}
\end{figure}

We use Q-residuals to examine the artifacts indicated in Figure~\ref{fig:results_longEval}c to Figure~\ref{fig:results_longEval}e. For each artifact, we project the participating atoms onto PC space with the dataset\del{,} and compute their Q-residuals to quantify where they are located in relation to their dataset. We hypothesize that these atoms are located in sparsely occupied regions where the model is forced to extrapolate, which would result in larger Q-residuals. Figure~\ref{fig:discussion_rxn_3ns_qresidual} illustrates the application of Q-residuals to the local atomic environments. We first isolate where the reaction begins, and find that elevated Q-residuals occur around 3.3 ns. We then project the atoms that participate in the deprotonation reaction from 3.3 to 3.8 ns, seen in Figure ~\ref{fig:results_longEval}c, onto PC space. 
    Figure~\ref{fig:discussion_rxn_3ns_qresidual} summarizes the first reaction in the 9 ns trajectory sampled by N-50pts. Figure~\ref{fig:discussion_rxn_3ns_qresidual}a is a PCA projection of the sampled environments from 3.3 ns to 3.8 ns and the fine-tuning dataset for N-50pts. The inlay represents an image taken from the simulation corresponding to the hydrogen environments sampled. Figure~\ref{fig:discussion_rxn_3ns_qresidual}b summarizes the Q-residual of each participating atom in the reaction of a \ce{Cl}, a \ce{H}, and an \ce{O}.  We observe that between 3.3 ns and 3.8 ns, which is highlighted in red, the Q-residual for \ce{H} increases to more than two times the standard deviation, indicating that the atomic environment for this atom is clearly an outlier. We relate this information to the PCA in Figure~\ref{fig:discussion_rxn_3ns_qresidual}a\add{,} where we project the sampled environments from 3.3 to 3.8 ns onto the dataset, and find that the hydrogen environments with elevated Q-residuals are outside of the dataset. This indicates that these specific environments are extrapolated by the model, and the reaction itself cannot be accurately evaluated. 

The residual then decreases after the hydrogen is deposited back onto another oxygen. This protonation is also unphysical as it protonates a carboxylic acid group, but Q-residuals decrease as the chemical environments \del{environment} (\ce{O-H}, \ce{C-OH}) are well within the model's datasets. Unfortunately, this highlights the fact that Q-residuals are strictly an outlier detector, and are not a measure \add{of} unphysical behavior in MD. Additionally, there is less movement in Q-residuals for oxygen or chlorine. This is likely because changes in atomic environments reflect smaller changes in SOAP descriptors, as indicated by the small regions occupied by other atom types in comparison to hydrogen, as demonstrated in Figure~\ref{fig:results_PCA_Datasets}, where the hydrogen datasets span larger regions of PCs than other elements. The limited span of other elements introduces another limitation in Q-residuals, as fluctuations in atomic environments will be less expressive than the hydrogen environments. 

\begin{figure}[H]
    \centering
    \includegraphics[width=1\linewidth]{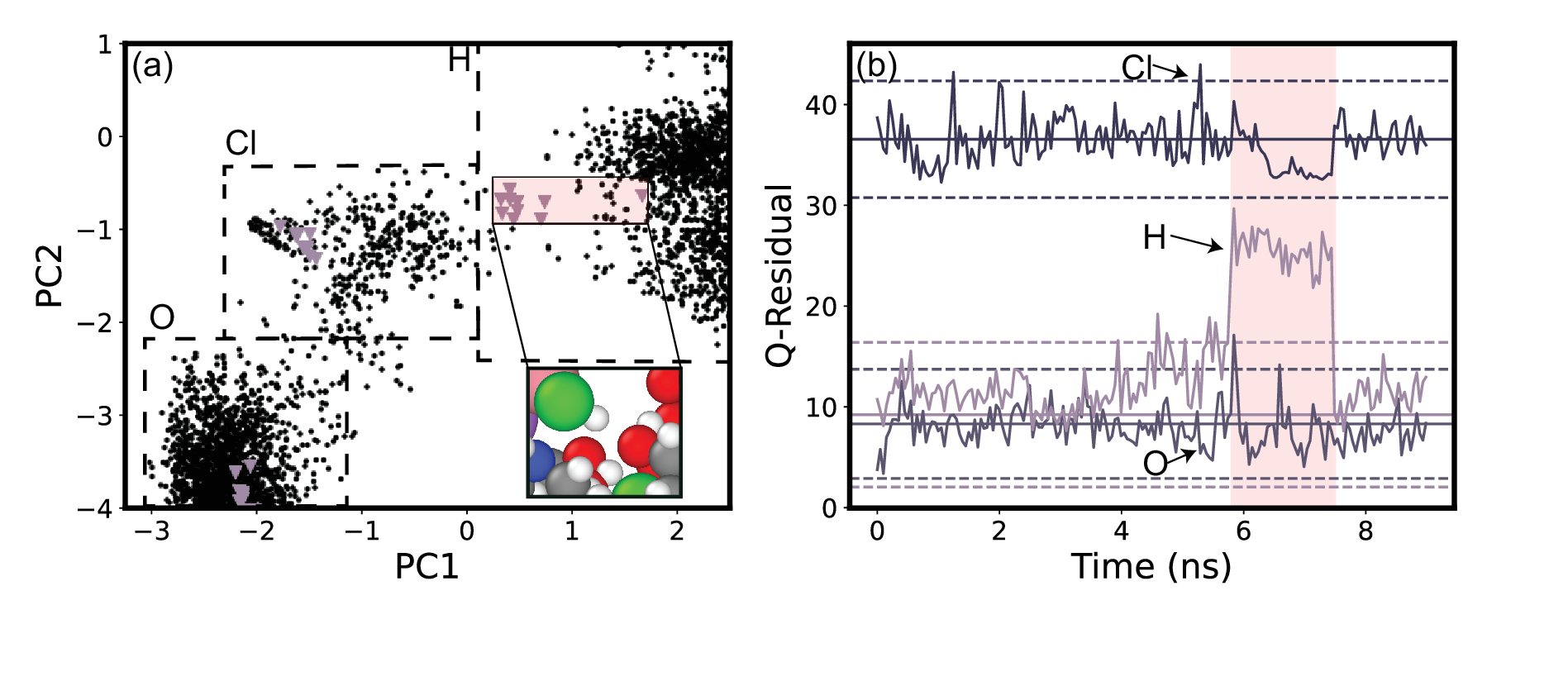}
    \caption{Another example of Q-residuals \add{(b)} indicating extrapolatory behavior\add{ in the PCA projection of SOAP descriptors (a) }during a 9-ns long N-50pts MD simulation between 5.8 and 7.5 ns.}
    \label{fig:discussion_rxn_5ns_qresidual}
\end{figure}

We then perform the same analysis for another deprotonation reaction, pictured in Figure~\ref{fig:results_longEval}d. Figure~\ref{fig:discussion_rxn_5ns_qresidual} displays the same PCA and Q-residual analysis as Figure~\ref{fig:discussion_rxn_3ns_qresidual}, but highlights a different deprotonation reaction that occurs 5.8 ns into the simulation.
Here, much like the deprotonation reaction shown in Figure~\ref{fig:discussion_rxn_3ns_qresidual}, we observe another region of elevated Q-residuals for the hydrogen atom in the reaction beyond two standard deviations \add{in Figure}~\ref{fig:discussion_rxn_5ns_qresidual}b, \add{that correlate to reactions and extrpolations in Figure}~\ref{fig:discussion_rxn_5ns_qresidual}a. Based on Figure~\ref{fig:discussion_rxn_3ns_qresidual} and Figure~\ref{fig:discussion_rxn_5ns_qresidual}, both deprotonation reactions that we observe occur when hydrogen explores a similar region outside of the dataset in PC space. 

Reactions like these deprotonation reactions are challenging to diagnose as they leave no noticeable impact on the energy evaluations, as seen in Figure~\ref{fig:results_longEval}c. The bond breaking reaction is captured within thermal vibrations of the liquid system at 300 K. Hence, through introducing Q-residuals, we have demonstrated a new way to capture and diagnose extrapolative behavior in \ce{H} atoms. In both cases, the deprotonation reactions represent \ce{H} environments that the model was not trained on. 

Even though the N-50 model is fine-tuned on 50 configurations, the local environments explored during the MD simulation are not available, causing the fine-tuned uMLIP to fail. In contrast, the fifty points that FT5, the \rep{iterative}{periodic} model, is fine-tuned, \rep{are}{is} almost fully representative of the \rep{bulk solvent}{9 ns trajectory}, as Figure~\ref{fig:results_timeResolvedPCA}b shows that the dataset (grey) fully covers the dynamics (plasma gradient). Furthermore, shown by Table~\ref{table:results_indTestSet}, although the \rep{iterative}{periodic} models are fine-tuned on data generated from just one trajectory, they perform better on an independently generated test set. We surmise that the difference is related to the training domains shown in Figure~\ref{fig:results_PCA_Datasets}c and Figure~\ref{fig:results_PCA_Datasets}d. While the naive dataset represents a wide, diffuse cloud of data in PC1 and PC2, it does not represent actual dynamics and \del{the} correlations between frames \rep{that}{are} occur in MD, evidenced by the significant missing coverage in Figure~\ref{fig:results_timeResolvedPCA}a. \add{We further investigate the consequences of these out-of-domain evaluations using TICA decompositions}\cite{molgedeySeparationMixtureIndependent1994} \add{in Figures}~\ref{fig:SI_TICA}~\add{and}~\ref{fig:SI_uMLIP_TICA}\add{, and find that out-of-domain evaluations lead to structural drifts over long timescales.}

In summary, we find that even when fine-tuned on system-specific structures of the same chemistry and composition, fine-tuned uMLIPs can still display errors that trace back to limitations in the fine-tuning dataset. Furthermore, we find that the uMLIP struggles to properly sample representative configurations of structures outside of its dataset, due to a known bias when sampling on out-of-domain structures. This means that iterative fine-tuning or advanced sampling techniques are necessary for fine-tuning, rather than simply synthesizing data with uMLIP-driven MD. 

We further apply our findings to another liquid system, composed of citric acid, lithium (I) chloride, and cobalt (II) chloride dissolved in water, and find that the \rep{iterative}{periodic} workflow indeed performs better than the naive models during MD (Figure ~\ref{fig:SI_DiluteSystem}). After training on 44 points, the naive workflow still exhibits a systematic error of +1.5 meV/at, whereas the \rep{iterative}{periodic} workflow demonstrated an error of 0.5 meV/at, centered around 0 meV/at. Additionally, as mentioned earlier, we also fine-tune on a 6.2-ns long (MD ran until 50 DFT points converged) trajectory sampled by the uMLIP, and report that such a workflow overfits the system. The result is that it performs well on MD, with errors of 3.37 meV/at (Table~\ref{table:SI_evalModelMetricsNoniterative}). However, this method performs poorly when evaluating a broad range of systems, resulting in an error of 12.32 meV/at on an independently generated test set (Table~\ref{table:SI_indTestSetNoniterative}). In summary, in either case, we show that biases inherent to the uMLIP negatively impact the downstream fine-tuned models.

A prominent feature of uMLIPs is their low, reported out-of-the-box errors, offering quantum accuracy on a large span of chemistries at minimal costs. Yet, in this paper, we suggest that uMLIPs must be fine-tuned several times to overcome an evident sampling bias. One possible simple sampling technique that may overcome these biases in liquid systems presented here is to generate system-specific structures using \del{a} software like Packmol, fine-tune on these structures, perform\del{ing} MD, and fine-tune on the following data\add{,} and repeat. Since the most improvement is seen after fine-tuning just once, we surmise that fine-tuning on system-specific structures, then fine-tuning on MD\add{,} may already allow for accurate evaluations. More advanced sampling techniques are an active field of research for MLIPs \cite{kulichenko_data_2024}, and may mitigate the effects of these biases in sampled datasets. Our study does not introduce new methods in this field; rather, we suggest that more advanced sampling techniques may be necessary to accurately generate data for fine-tuning. We find that simply fine-tuning the model on self-sampled data is insufficient for configurational diversity\del{,} and results in unphysical models.

In terms of high-throughput screening, this unfortunately suggests that system-specific models need to be fine-tuned on more than just data generated from a universal potential. This aligns with the findings of Chorna et al., who show that fine-tuned potentials preserve the latent representations learned during training, thereby retaining biases encountered while training  \cite{chorna_comparing_2025}. Thus, any biases encountered in a case like our naive fine-tuning approach will be \rep{propagated}{propogated} through fine-tuning. This bias is consistent with the unphysicalities observed in MD, which appeared to be largely independent of the number of fine-tuning data points. \rep{Iteratively}{Periodically} fine-tuning may sidestep this effect by iteratively reducing bias through consecutively sampled data. 

\subsection{Limitations \& Future Work}
In this work, we use \verb|MACE-MP-0b|, and do not investigate other model architectures. However, systematic biases have been observed for models such as CHGNET and M3GNet as well \cite{deng_systematic_2025}, and we expect the above findings to generalize to these other models \cite{hanserothFineTuningUnifiesFoundational2026}. We also expect these findings to generalize to other uMLIPs, since out-of-distribution errors are a major limitation of MLIPs and uMLIPs \cite{kreiman_understanding_2026}. This work demonstrates that out-of-distribution errors necessitate careful consideration of the fine-tuning dataset. 

Additionally, we notice that forces on chlorines are systematically underpredicted for the \rep{iterative}{periodic} models, while the naive models are more accurate in this case, as demonstrated in supplementary material (Figure~\ref{fig:SI_Cl_Forces}). \add{This specifically happens to the Cl atoms that are coordinated with Co, forming a LiCoCl$_3$ cluster stable for 7 nanoseconds. Such a LiCoCl3 configuration does not appear in the FT5 fine-tuning dataset, reminiscent of a misprediction already observed in the naïve workflow, where we observe that the under-coverage of local environments that are vital for production MD, which results in spurious reactions and model failures. We discuss this misprediction in more detail in Supplemental Information, with Figure}~\ref{fig:SI_violin} \add{ demonstrating the error distribution, with high-error chlorines being those within LiCoCl$_3$}. 
\rep{This highlights the need for both data diversity as well as active learning iterations, and is a disadvantage of only one starting configuration, especially for systems with slow dynamics. Restarting from an independent image after each fine-tuning iteration may combine the data diversity advantage from naive fine-tuning and the realistic structural diversity of iterative fine-tuning.}{This lack of data highlights the need for both data diversity and active learning iterations, and is a disadvantage of using only one starting configuration, especially for systems with slow dynamics.} Thus, we suggest hybrid approaches where one may train on multiple independent trajectories, with at least two fine-tuning steps. \add{We show that fine-tuning iteratively can reduce the energetic errors and force variance. To obtain accuracy on material properties sensitive to the force errors (e.g., thermal conductivity} \cite{wuCorrectingForceErrorinduced2024}\add{), more diverse data sampled through hybrid strategies as discussed earlier and fine-tuning steps are necessary.}

Sampling high forces using MD is rare, and we do not expect our models to perform well at high forces. To sample at high forces, one may distort atoms, run MD at high temperatures, or generate high-energy structures using Packmol. The methods presented are simple, practical strategies. They reflect an approach that \add{a} typical user might take when fine-tuning a model on a new material. These methods may not be the most efficient at sampling chemical space, but they demonstrate the limitations of uMLIPs for data generation. Regardless, our strategies may produce overlapping and redundant data as we select data for retraining at fixed intervals. To develop a dataset that is compatible \rep{with}{for} different conditions, one may develop a more efficient dataset generation strategy by altering temperature \cite{pitfield_active_2025}, or running multiple trajectories in parallel for systems of different temperature, pressure, or composition. However, we expect our findings to generalize to other systems, as evidenced by our comparisons with the aqueous system.

\section*{Conclusion}

In this work, a universal MACE machine-learned interatomic potential, \verb|MACE-MP-0b| is fine-tuned using two different dataset generation techniques aimed at isolating and evaluating bias in universal potentials. The system is completely different from the training dataset, \rep{consisting}{comprised} of a liquid system \add{of} choline-chloride citric acid with dissolved lithium (I) and cobalt (II) ions. We find a persistent bias in using universal potentials in the sampling of configurations that results in flexible bonds near-equilibrium, which we demonstrate for \ce{O-H} bonds. Using these configurations solely to fine-tune results in problematic downstream effects such as \rep{fictitious}{fictious} reactions. Our findings indicate that such unphysical behavior is a result of extrapolation, where configurations encountered during MD with fine-tuned models \rep{are}{is} not encountered by MD with uMLIPs. We report that it is necessary to perform an \rep{iterative workflow}{active learning loop} to effectively represent a system's configuration space and generate a dataset for MD. Running simulations for longer using universal potentials results in overfit models. In summary, uMLIPs have direct caveats that negatively impact their out-of-the-box versatility. Fine-tuning on more data does not always correlate with more accurate models, as we have shown, and synthesizing datasets needs to be a careful process.
\section*{Methods}
\subsection{System Definition}
We follow the system presented by Peeters et al. Each system contains a 1:2 molar ratio of \ce{[(CH3)3NCH2CH2OH]+Cl^-} (choline chloride) to \ce{C6H8O7} (citric acid), along with \ce{LiCl} and \ce{CoCl2} salts \cite{peeters_solvometallurgical_2020}. Starting configurations were generated using Packmol \cite{martinez_packmol_2009}. \add{Each random structure was generated enforcing a cubic simulation cell with a side length of 14\AA, comprised of 1 Li, 1 Co, 6 Cl, 3 choline, and 6 citric acid molecules. A minimum interatomic tolerance of 1.5\AA  was imposed, and the movebadrandom setting was enabled, which moves poorly placed molecules to new random positions and assists with convergence.}, resulting in a density of 1.14 g/cm\textsuperscript{3}, which is the room temperature density of \ce{LiCl} and \ce{CoCl2}, and ChCl:CA. The final systems contain 197 atoms.

\subsection{Molecular Dynamics}
Molecular dynamics (MD) were performed using LAMMPS molecular dynamics package \cite{thompson_lammps_2022}. All simulations were run in the isothermal-isobaric (NPT) ensemble for 1.0 ns at 0 bar, 300 K\add{,} using timesteps of 0.5 fs. Temperature and pressure were controlled using a Nose-Hoover thermostat and barostat with damping parameters of 50 fs and 500 fs\add{,} respectively. Within each simulation, after 0.5ns of equilibration, we select 11 images for evaluation (equal intervals of 50 picoseconds).

\subsection{Model}
We utilized the MACE foundation model \verb|MACE-MP-0b/small| for MD \cite{batatia_foundation_2024}. We found that this model was the best trade-off between accuracy and speed. 

\subsection{DFT}
We computed single\add{-}point density functional theory (DFT) using the Vienna Ab initio Simulation Package (VASP) \cite{kresse_ab_1993, kresse_efficiency_1996, noauthor_efficient_nodate}. DFT was employed using the Projector-Augmented-Wave method and Perdew-Burke-Ernzerhof (PAW-PBE) exchange-correlation functional \cite{perdew_generalized_1996, kresse_ultrasoft_1999}. Hubbard $U$ corrections were computed for \ce{CoCl2} using linear response theory resulting in $U_{Co}=5.76$ eV (Figure~\ref{fig:SI_LinearResponse})\cite{cococcioni_linear_2005}. Gaussian smearing with $\sigma=0.1$ was applied, and electronic convergence (EDIFF) was set to 10$^{-5}$ eV.

\subsection{Fine-tuning}
Multihead fine-tuning is a training strategy that allows evaluations to be made for different DFT approximations and other first-principles calculations using the same model \cite{batatia_foundation_2024}. During fine-tuning\add{,} a subset of the initial training data is kept and re-trained on to prevent catastrophic forgetting \cite{kirkpatrick_overcoming_2017}. Presently, we randomly select a subset of 30,000 structures from the universal potential's training set as a replay set. Fine\add{-}tuning dataset structures were generated with the procedures discussed below, and were weighted at a 10:1 ratio (suggested by MACE developers \cite{noauthor_multihead_nodate}), favoring the fine tuning dataset. 

\subsubsection{Hyperparameter Selection}
Hyperparameters were chosen to emulate training the foundation model with a lower learning rate to accommodate the smaller datasets \cite{batatia_foundation_2024}: \textsubscript{E}=1, w\textsubscript{F}=10, w\textsubscript{S}=10, lr=10\textsuperscript{-4}. We choose an 80/20 train test split. Fine-tuning was allowed to run for 250 epochs or until no improvement was seen for 40 epochs. 

\subsection{Model Evaluation \& Metrics}
We use the \verb|dscribe| python package \cite{laakso_updates_2023} to decompose datasets into smooth-overlap-of-atomic-positions (SOAP) parameters to represent chemical environments, and principal component analysis (PCA) to visualize the span of each dataset. We determined the SOAP parameters using the following hyperparameters: $r_{cut} = 5$, $n_{max} = 8$, $\ell_{max} = 6$, $\sigma = 0.375$. For each PCA, we use 5 principal components, and the variance captured by the PCs is 68.79\%.

To evaluate each model (i.e.\add{,} N-10pts, N-21pts, FT1, FT2, etc.) sample a 1 ns trajectory from the same independent starting configuration and evaluate the production run through root mean squared errors (RMSEs) in potential energies, atomic forces, and stresses as computed in methods.
\begin{equation}
    \text{RMSE}_E = \sqrt{\frac{1}{N}\sum_{i=1}^N(E_i^{DFT}-E_i^{PRED})^2} \label{eq:rmse_energy}
\end{equation}
\begin{equation}
    \text{RMSE}_{F_\theta} = \sqrt{\frac{1}{N}\sum_{i=1}^N(F_{i\theta}^{DFT}-F_{i\theta}^{PRED})^2} \label{eq:rmse_force}
\end{equation}
\begin{equation}
    \text{RMSE}_{\sigma} = \sqrt{\frac{1}{6N}\sum_{i=1}^N\sum_{\alpha\leq\beta}(\sigma_{\alpha\beta}^{DFT}-\sigma_{\alpha\beta}^{PRED})^2} \label{eq:rmse_stress}
\end{equation}
\begin{equation}
    \text{Rel. F Err} = \frac{\text{RMSE}_{F}}{\dfrac{1}{3N}\sum_{i=1}^{N}\sum_{\theta \in \{x,y,z\}}|F_{i\theta}^{DFT}|} \times 100\% \label{eq:relerror}
\end{equation}
\subsection{Q-Residuals}
To quantify how far a trajectory deviates from its dataset, we first construct a SOAP/PCA using the dataset. We next project the trajectory onto the new basis set. Let the dataset be
\begin{itemize}
    \item $\mathbf{X} \in \mathbb{R}^{n \times p}$: dataset of $n$ samples with $p$ features
    \item $\boldsymbol{\mu} \in \mathbb{R}^{1 \times p}$: feature-wise mean vector
    \item $\mathbf{W} \in \mathbb{R}^{p \times k}$: PCA weight matrix (first $k$ components)
    \item $\mathbf{Z} \in \mathbb{R}^{n \times k}$: PCA projection matrix
\end{itemize}
where $n$ is the number of samples ($n_{frames}\times n_{atoms}$), $p$ is the number of SOAP features, $k$ is the number of PCA features. PCA projection is a linear transformation described as
\begin{equation*}
    \mathbf{Z} = (\mathbf{X}-\boldsymbol{\mu})\cdot \mathbf{W}
\end{equation*}
We then reconstruct the projected vector back into its original form,
\begin{equation*}
    \boldsymbol{\hat X} = \mathbf{Z}\mathbf{W}^T+\boldsymbol{\mu}
\end{equation*}
The squared residual error is then
\begin{equation*}
    \mathbf{E} = (\mathbf{X}-\boldsymbol{\hat X})(\boldsymbol{\hat X}-\mathbf{X})
\end{equation*}
The Q-residual (squared prediction error) for each sample is computed as the row-wise sum of squared residuals
\begin{equation*}
    Q_i = \sum_{j=1}^p E_{ij}
\end{equation*}
Q-residuals describe the variation lost through PCA. In essence, they describe how well the trajectories align with their reference datasets. Larger Q values indicate points that are far from reference data, and smaller Q values indicate points that are close to reference data.  
\section*{Supporting Information}
\add{Supporting Information: Hubbard U correction linear response, metrics for noniterative workflow, aqueous system results, bond-length histograms for CO and HCl, cobalt coordination analysis, per-element force metrics, and TICA analysis.}
\section*{Data Availability}
All data used in this study is available on GitHub at \url{https://github.com/YangLab-GT/umlip_bias_finetuning}
\section*{Acknowledgements}
N.W. acknowledges that this work used Delta GPU at NCSA through allocation CHM250008 from the Advanced Cyberinfrastructure Coordination Ecosystem: Services \& Support (ACCESS) program, which is supported by U.S. National Science Foundation grants \#2138259, \#2138286, \#2138307, \#2137603, and \#2138296. N.W. acknowledges that this work was supported in part through research cyberinfrastructure resources and services provided by the Partnership for an Advanced Computing Environment (PACE) at the Georgia Institute of Technology, Atlanta, Georgia, USA. RRID:SCR\_027619. J.Y. acknowledges support from startup funds at the Georgia Institute of Technology School of Chemical and Biomolecular Engineering, the Pirkle Faculty Fellowship, and Google. 
\bibliography{uMLIP_manuscript_bibliography_jctc}
\end{document}